\documentclass[11pt,twoside]{article} % Specifies the document style.
\usepackage{times,fancyhdr}
\usepackage{cite}
\usepackage{graphicx}

\date{February 21, 2007}

\title{Superconductors with Topological Order and their Realization in Josephson
Junction Arrays} % Declares the document's title.
\author{M. Cristina Diamantini, Pasquale Sodano and Carlo A. Trugenberger} % Declares the author's name.

\begin{document} % End of preamble and beginning of text.

\pagestyle{fancy}
\fancyhead{} % clear all header fields
\fancyhead[EC]{M. Cristina Diamantini, Pasquale Sodano and Carlo A. Trugenberger}
\fancyhead[EL,OR]{\thepage}
\fancyhead[OC]{Superconductors with Topological Order and their Realization in Josephson
Junction Arrays}
\fancyfoot{} % clear all footer fields
\renewcommand\headrulewidth{0.5pt}
\addtolength{\headheight}{2pt} % make space for the rule

\maketitle % Produces the title.
We will describe  a new superconductivity mechanism, proposed by the authors in
\cite{nsm},
 which is based on a topologically ordered 
ground state rather than on the usual Landau mechanism of spontaneous symmetry
 breaking. Contrary to anyon superconductivity it works in any dimension
and it preserves P-and T-invariance. In particular we will discuss the 
low-energy effective field theory, what would be the Landau-Ginzburg
formulation for conventional superconductors.

\section{Introduction}

For many years the theory of phase transitions was entirely understood in
terms of the Landau-Ginzburg theory, based on symmetry breaking, order parameters and on the
mathematical framework of group theory. With the discovery of the fractional
quantum Hall liquids \cite{Lau}, that are incompressible and exist only at some "magical"
filling fractions, it was understood that the internal order characterizing
these states is a new type of order, different from any other known type of order. This new type of
order is the topological order \cite{wen}, a particular type of quantum order.

Quantum order describes the zero-temperature properties of a quantum ground state, and
characterizes universality classes of quantum states, described by {\it complex}
ground state wave-functions \cite{wen}. Quantum phase transition are characterized by
changes in the quantum entanglement properties of these complex ground state
wave-functions.
Topological order is a special type of quantum order whose hallmarks are 
the presence of a gap for all excitations (incompressibility) 
and the degeneracy of the ground state on manifolds
with non-trivial topology \cite{wen}. In the case of the fractional quantum Hall effect,
which is a (2+1)-dimensional system, another hallmark
is the presence of excitations with fractional spin and statistics, 
called anyons \cite{wilczek}. The long-distance properties
of these topological fluids can be explained by an infinite-dimensional $W_{1+\infty}$
dynamical symmetry \cite{czt}, and are described by effective Chern-Simons field theories
 \cite{wen2} with compact gauge group, which break P-and 
T-invariance. 

The topologically ordered superconductors we propose have a long-distance 
hydrodynamic action which 
can be entirely formulated in terms of generalized compact gauge fields, the dominant 
term being the topological BF action \cite{birmi}.  The BF
theory has many applications, from 2D gravity \cite{marito} to the mathematical characterization of embedded
manifolds \cite{gord}.
It turns out that it is also the general model that describes the long-distance behaviour  
of systems that exhibit P- and T-preserving
topological order in any dimension. It reduces to a doubled Chern-Simons
model in (2+1)-dimensions \cite{freedman}.
The existence of such non-conventional superconductors
is also supported by purely algebraic considerations \cite{carlo}.

In \cite{dst}, we have proven that planar Josephson junction arrays (JJA)
can be exactly mapped onto an Abelian gauge theory
with a mixed Chern-Simons term (BF-model): JJA provide thus a first concrete example of 
topological superconductors. The Abelian gauge theory exactly reproduces the phase diagram of
JJA and the insulator/superconductor quantum phase transition at $T=0$ \cite{jja}.
JJA  have also been recently  considered by several other
authors \cite{as, joffe},  as
controllable devices which  exhibit topological order. 

The idea that gauge fields can be used to model the long distance behaviour of
condensed matter systems was originally proposed in\cite{froa},
and particularly exploited for planar systems \cite{fwz}.
In a nutshell, the idea is that charge fluctuations around a given
ground state are decribed by a conserved current $j^{\mu }$, which in (2+1)
dimensions can be represented in terms of a gauge field  $B_{\mu}$ according
to $j^{\mu } \propto \epsilon ^{\mu \alpha \nu}\partial _{\alpha }B_{\nu}$.
For a wide
class of systems the effective action governing the dynamics of the charge
fluctuations is quadratic in the gauge fields $B_{\mu }$ at long distances
\cite{froa} . Clearly this effective action must also be gauge invariant, reflecting the
original gauge invariance of the current $j^{\mu }$: one
obtains thus an effective gauge theory at long distances (which is not
necessarily relativistic).
The ground states of a wide class of planar condensed matter systems
\cite{fra}
can be classified according to the lowest derivative terms appearing in
their effective
gauge theories at long distances. This way, the Chern-Simons term describes
incompressible quantum fluids (quantum Hall states) and chiral spin liquids
\cite{frob} while the Maxwell term describes a
(2-dim.) superfluid (superconductor).

The case of JJA is different in the sense that the gauge theory description  is an exact
mapping and not only an effective theory \cite{dst}: JJA can be {\it exactly mapped} onto an Abelian gauge
theory with two gauge fields describing a current of charges and a current of
vortices coupled by a mixed Chern-Simons term:
\begin{equation}{\cal L}=-{1\over 4e^2}F_{\mu \nu}F^{\mu \nu } + {\kappa \over 2\pi }
A_{\mu }\epsilon ^{\mu \alpha \nu }\partial_{\alpha }B_{\nu }-{1\over 4g^2}
f_{\mu \nu }f^{\mu \nu } \ ,
\label{mcs}
\end{equation}
Global superconductivity in planar JJA is thus the simplest example of
the new mechanism of superconductivity we propose. 

We argue \cite{bff} that  a less trivial example are frustrated JJA:
these may support a
topologically ordered superconducting ground state characterized by
a non-trivial ground state degeneracy on the torus. 
These superconducting quantum fluids provide explicit examples of 
systems in which superconductivity arises from a topological
mechanism rather than from the usual Landau-Ginzburg mechanism.

The paper is organized as follows. In Section 2 we will review the definition
of the Chern-Simons model on the lattice. In Section 3 we show how planar
JJA can be exactly mapped onto a gauge theory with two
Maxwell fields coupled by a mixed, periodic, Chern-Simons term. Section 4 is
devoted to the study of the phase diagram, showing how the
superconductor/insulating transition of the JJA is reproduced
by the gauge theory. In Section 5 we present the general theory of topological
superconductors in any space-time dimension. Section 6 is, instead, devoted to
the study of the role of frustration.

\section{Lattice Chern-Simons model}
Our model (\ref{mcs}) \ can be rewritten in terms of the dual field strengths
\begin{eqnarray}F^{\mu } &&\equiv {1\over 2}\epsilon^{\mu \alpha \beta}
F_{\alpha \beta}\ ,\qquad \qquad F_{\mu \nu} \equiv \partial _{\mu }A_{\nu }
-\partial _{\nu }A_{\mu } \ ,\nonumber \\
f^{\mu } &&\equiv {1\over 2} \epsilon ^{\mu \alpha \beta} f_{\alpha \beta}
\ ,\qquad \qquad f_{\mu \nu} \equiv \partial _{\mu }B_{\nu }-\partial _{\nu }
B_{\mu } \ , \
\label{dfs}
\end{eqnarray}
as follows (throughout this paper we use units such that
$c=1$ and $\hbar =1$.)
\begin{equation}{\cal L}_{CS} = -{1\over 2e^2} \left({1\over \eta}F_0F^0
+F_iF^i\right)
+{\kappa \over 2\pi }
A_{\mu }\epsilon ^{\mu \alpha \nu } \partial _{\alpha }B_{\nu } -
{1\over 2g^2}
\left( {1\over \eta }f_0f^0 +f_if^i\right) \ .
\label{mod}
\end{equation}
For later convenience we have introduced a magnetic permeability $\eta $,
equal for the two gauge fields. The coupling constants $e^2$ and $g^2$ have
dimension mass, whereas  the coefficient $\kappa $ of the mixed Chern-Simons
term is dimensionless. Note that we take $B_{\mu }$ to represent a
pseudovector
gauge field, so that the mixed Chern-Simons term does not break the discrete
symmetries of parity and time reversal.

The action corresponding to (\ref{mod}) \ is separately invariant under the two
Abelian gauge transformations
\begin{eqnarray}A_{\mu } &&\to A_{\mu } +\partial _{\mu }\lambda \ ,\nonumber \\
B_{\mu } &&\to B_{\mu } +\partial _{\mu }\omega \ ,\
\label{agt}
\end{eqnarray}
with gauge groups $R_A$ and $R_B$, respectively. Moreover, the action is
also invariant under the {\it duality transformation}
\begin{eqnarray}A_{\mu } &&\leftrightarrow B_{\mu }\ ,\nonumber \\
e &&\leftrightarrow g\ ,\
\label{dtr}
\end{eqnarray}
so that the model is {\it self-dual}.

The Lagrangian (\ref{mod})  \ can be easily diagonalized by the linear transformation
\begin{eqnarray}A_{\mu } &&= \sqrt {e\over g} \left( a_{\mu }+b_{\mu }
\right)
\ ,\nonumber \\
B_{\mu } &&= \sqrt{g\over e} \left(a_{\mu }-b_{\mu } \right) \ .\ 
\label{lit}
\end{eqnarray}
In terms of these  new variables the model (\ref{mod})  \ describes a free theory,
\begin{equation}{\cal L}_{CS}= -{1\over eg} \left( {1\over \eta }G_0G^0
+G_iG^i \right)
+{\kappa \over 2\pi }a_{\mu }\epsilon^{\mu \alpha \nu }\partial _{\alpha}
a_{\nu } - {1\over eg} \left({1\over \eta }g_0g^0 +g_ig^i\right) -
{\kappa \over 2\pi}b_{\mu }\epsilon^{\mu \alpha \nu}\partial_{\alpha }b_{\nu }
\ ,
\label{tla}
\end{equation}
where $G^{\mu }$ and $g^{\mu }$ are the dual field strengths for the new gauge
fields $a_{\mu }$ and $b_{\mu }$, respectively. This Lagrangian describes
a doublet of excitations with topological mass \cite{jac}
\begin{equation}m={|\kappa|eg\over 2\pi }\ ,
\label{tma}
\end{equation}
and  spectrum
\begin{equation}E({\bf q})=\sqrt{m^2+{1\over \eta}|{\bf q}|^2} \ .
\label{spe}
\end{equation}

In the following we shall formulate a Euclidean lattice version of
the above Chern-Simons model. To this end we introduce a three-dimensional
rectangular lattice with lattice spacings $l_{\mu }$ in the three directions.
In particular we shall take the lattice spacings $l_1=l_2\equiv l$ and
identify $l_0$ with the spacing in the Euclidean time direction. Lattice
sites are denoted by the three-dimensional vector $x$; the gauge fields
$A_{\mu }(x)$ and $B_{\mu }(x)$ are associated with the links $(x, \mu )$
between the sites $x$ and $x+\hat \mu $, where $\hat \mu $ denotes a unit
vector in direction $\mu $ on the lattice.

On the lattice we introduce the following forward and backward derivatives and
shift operators:
\begin{eqnarray}d_{\mu } f(x) &&\equiv {{f(x+l_{\mu }\hat \mu )-f(x)}\over
l_{\mu }}\ ,\qquad \qquad S_{\mu }f(x) \equiv f(x+l_{\mu }\hat \mu )\ ,\nonumber \\
\hat d_{\mu } f(x) &&\equiv {{f(x)-f(x-l_{\mu }\hat \mu )}\over l_{\mu }} \ ,
\qquad \qquad \hat S_{\mu }f(x) \equiv f(x-l_{\mu }\hat \mu ) \ . \
\label{dso}
\end{eqnarray}
Summation by parts on the lattice interchanges both the two derivatives (with
a minus sign) and the two shift operators; gauge transformations are defined
using the forward lattice derivative. Corresponding to the two derivatives
in \ref{dso} , we can define also two lattice analogues of the Chern-Simons
operators
$\epsilon _{\mu \alpha \nu }\partial _{\alpha }$
\cite{frod}\ \cite{dst} :
\begin{equation}k_{\mu \nu } \equiv S_{\mu } \epsilon _{\mu \alpha \nu } d_{\alpha }
\ , \qquad \qquad \hat k_{\mu \nu } \equiv \epsilon _{\mu \alpha \nu } \hat d
_{\alpha }\hat S_{\nu } \ ,
\label{lcs}
\end{equation}
where no summation is implied over equal indices $\mu $ and $\nu $.
Summation by parts on the lattice interchanges also these two operators
(without an extra minus sign).
The operators \ref{lcs} \ are both local and gauge invariant, in the sense that
\begin{equation}k_{\mu \nu} d_{\nu }=\hat d_{\mu }k_{\mu \nu }=0\ ,\qquad \qquad
\hat k_{\mu \nu }d_{\nu }=\hat d_{\mu }\hat k_{\mu \nu } =0 \ ,
\label{gin}
\end{equation}
and their product reproduces the relativistic, Euclidean lattice
Maxwell operator:
\begin{equation}k_{\mu \alpha }\hat k_{\alpha \nu } = \hat k_{\mu \alpha }
k_{\alpha \nu } = -\delta _{\mu \nu }\nabla ^2 +d_{\mu }\hat d_{\nu } \ ,
\label{lmo}
\end{equation}
where $\nabla ^2 \equiv \hat d_{\mu }d_{\mu }$ is the
three-dimensional Laplace
operator. Using $k_{\mu \nu }$ we can also define the lattice dual field
strengths as
\begin{eqnarray}F_{\mu } &&\equiv \hat k_{\mu \nu }A_{\nu }\ ,\nonumber \\
f_{\mu } &&\equiv k_{\mu \nu }B_{\nu } \ .\
\label{ldfs}
\end{eqnarray}
The identity (\ref{lmo}) \ then tells us that we can simply write the relativistic,
Euclidean lattice Maxwell terms as $\sum _x F_{\mu }F_{\mu }$ and
$\sum _x f_{\mu }f_{\mu }$.

Using all these definitions we can now write the Euclidean lattice partition
function of our model (\ref{mod})  \ as follows:
\begin{eqnarray}Z_{CS} &&= \int {\cal D}A_{\mu }
\int {\cal D}B_{\mu } \ {\rm exp}(-S_{CS}) \ ,\nonumber \\
S_{CS} &&= \sum_x {l_0l^2\over 2e^2} \left( {1\over \eta }F_0F_0
+F_iF_i \right)
-i {l_0l^2\kappa \over 2\pi } A_{\mu }k_{\mu \nu }B_{\nu } + {l_0l^2 \over
2g^2} \left( {1\over \eta }f_0f_0 +f_if_i \right) \ ,\
\label{lpf}
\end{eqnarray}
where we have introduced the notation ${\cal D}A_{\mu }\equiv
\prod _{(x, \mu )} dA_{\mu }(x)$ and gauge fixing is understood.

For later convenience we introduce also the finite difference operators
\begin{equation}\Delta _{\mu } \equiv l_{\mu }d_{\mu }\ , \qquad \qquad
\hat \Delta _{\mu } \equiv l_{\mu }\hat d_{\mu } \ ,
\label{fdo}
\end{equation}
where no summation over equal indices is implied. Correspondingly, we
introduce also the finite difference analogue of the operators $k_{\mu \nu }$
and $\hat k_{\mu \nu }$:
\begin{equation}K_{\mu \nu } \equiv S_{\mu }\epsilon _{\mu \alpha \nu }
\Delta _{\alpha } \ ,\qquad \qquad \hat K_{\mu \nu } \equiv \epsilon_{\mu
\alpha \nu }\hat \Delta_{\alpha }\hat S_{\nu } \ .
\label{kbig}
\end{equation}
These satisfy equations analogous to (\ref{gin}) \ and (\ref{lmo}) \ with all derivatives
substituted by finite differences.

\section{Josephson junction arrays}

JJA we analize are quadratic, planar arrays of spacing
$l$ of superconducting islands with nearest neighbours Josephson couplings
of strength $E_J$. Each island has a capacitance $C_0$ to the ground;
moreover there are also nearest neighbours capacitances $C$. The Hamiltonian
characterizing such systems is thus given by
\begin{equation}H=\sum_{\bf x} \ {C_0\over 2} V_{\bf x} + \sum _{<{\bf x \bf y}>}
\left( {C\over 2} \left( V_{\bf y}-V_{\bf x} \right) ^2 + E_J
\left( 1-{\rm cos}\ N\left( \Phi _{\bf y} - \Phi _{\bf x} \right) \right)
\right) \ ,
\label{hjja}
\end{equation}
where boldface characters denote the sites of the two-dimensional array,
$<{\bf x \bf y}>$ indicates nearest neighbours, $V_{\bf x}$ is the electric
potential of the island at ${\bf x}$ and $\Phi _{\bf x}$ the phase of its
order parameter.  For generality we allow for any integer $N$ in the Josephson
coupling, so that the phase has periodicity $2\pi /N$: obviously $N=2$ for the
real systems.

With the notation introduced in the previous section the
Hamiltonian (\ref{hjja}) \ can be rewritten as
\begin{equation}H= \sum_{\bf x} \ {1\over 2} V \left( C_0 - C\Delta \right) V +
\sum _{{\bf x},i} \ E_J \left( 1-{\rm cos} \ N \left( \Delta_i \Phi \right)
\right) \ ,
\label{hjjb}
\end{equation}
where $\Delta \equiv \hat \Delta_i \Delta_i$ is the two-dimensional finite
difference Laplacian and we have omitted the explicit location indeces on
the variables $V$ and $\Phi$.

The phases $\Phi _{\bf x}$ are quantum-mechanically conjugated to the charges
$Q_{\bf x}$ on the islands: these are quantized in integer multiples of $N$
(Cooper pairs for $N=2$):
\begin{equation}Q= q_e N p_0\ ,\qquad \qquad p_0 \in Z\ 
\label{cqu}
\end{equation}
where $q_e$ is the electron charge. The Hamiltonian (\ref{hjjb}) \ can be expressed
in terms of charges and phases by noting that the electric
potentials $V_{\bf x}$ are determined by the charges $Q_{\bf x}$ via
a discrete version of Poisson's equation:
\begin{equation}\left( C_0 -C\Delta \right) V_{\bf x} = Q_{\bf x} \ .
\label{dvp}
\end{equation}
Using this in (\ref{hjjb}) \  we get
\begin{equation}H= \sum_{\bf x} \ N^2E_C \ p_0 {1\over {C_0\over C}-\Delta } p_0 +
\sum _{{\bf x},i} \ E_J \left( 1-{\rm cos} \ N\left( \Delta _i \Phi \right)
\right) \ ,
\label{hjjc}
\end{equation}
where $E_C\equiv q_e^2/2C$. The integer charges $p_0$ interact via a
two-dimensional Yukawa potential of mass $\sqrt{C_0/C} /l$. In the
nearest-neighbours capacitance limit $C\gg C_0$, which is accessible
experimentally, this becomes essentially a two-dimensional Coulomb law.
>From now on we shall consider the limiting case $C_0=0$. In this case the
{\it charging energy} $E_C$ and the {\it Josephson coupling} $E_J$ are the
two relevant energy scales in the problem. These two massive parameters can
also be traded for one massive parameter $\sqrt{2N^2E_CE_J}$, which represents
the {\it Josephson plasma frequency} and one massless parameter $E_J/E_C$.

The zero-temperature partition function of the Josephson junction array
admits a (phase-space) path-integral representation \cite{nor}. Since the variables $p_0$ are integers, the
imaginary-time integration has to be performed stepwise; we
introduce therefore
a lattice spacing $l_0$ also in the imaginary-time direction. This has to be
just smaller of the typical time scale
on which the integers $p_0$ vary, in
the present case the inverse of the Josephson plasma frequency:
$l_0 \le O\left( 1/\sqrt{2N^2E_CE_J} \right) $. We thus get the following
partition function:
\begin{eqnarray}Z &&= \sum_{\{p_0\}} \int_{-\pi /N}^{+\pi /N} {\cal D}\Phi
\ {\rm exp}(-S)\ ,\nonumber \\
S &&= \sum _x -iN\ p_0\Delta _0\Phi + N^2 E_C l_0 \ p_0 {1\over -\Delta } p_0
+\sum _{x, i} \ l_0E_J \left( 1- {\rm cos}\ N \left( \Delta _i \Phi \right)
\right) \ ,\
\label{pfjja}
\end{eqnarray}
where now the sum in the action $S$ extends over the three-dimensional lattice
with spacing $l_0$ in the imaginary time direction and $l$ in the spatial
directions.

In the next step we introduce vortex degrees of freedom by replacing the
Josephson term by its Villain form:
\begin{eqnarray}Z &&= \sum _{{\{p_0\} \atop \{v_i\}}}
\int _{-\pi /N}^{+\pi /N} {\cal D}\Phi \ {\rm exp}(-S)\ ,\nonumber \\
S &&= \sum_x -iN\ p_0\Delta _0 \Phi + N^2 E_C l_0 \ p_0 {1\over -\Delta } p_0
+N^2 l_0 {E_J\over 2} \left( \Delta _i \Phi + {2\pi \over N} v_i \right) ^2
\ .\
\label{pfjjb}
\end{eqnarray}
Strictly speaking, this substitution is valid only for $l_0 E_J \gg1$; however
the Villain approximation retains all most relevant features of the
Josephson coupling for the whole range of values of the coupling
$E_J$  and therefore we shall henceforth adopt it.

We now represent the Villain term as a Gaussian integral over real variables
$p_i$ and we transform also $p_0$ to a real variable by introducing new
integers $v_0$ via the Poisson summation formula
\begin{equation}\sum_{k=-\infty }^{k=+\infty } {\rm exp}(i2\pi kz)= \sum_{n=-\infty}
^{n=+\infty } \delta (z-n) \ .
\label{poi}
\end{equation}
By grouping together the real and integer $p$ and $v$ variables into
three-vectors $p_{\mu }$ and $v_{\mu }$, $\mu=0,1,2$ we can write the
partition function as
\begin{eqnarray}Z &&= \sum _{\{ v_{\mu } \} } \int
{\cal D}p_{\mu } \int_{-\pi /N}^{+\pi /N} {\cal D}\Phi \ {\rm exp}(-S)\ ,\nonumber \\
S &&= \sum_x -iN p_{\mu } \left( \Delta _{\mu } \Phi + {2\pi \over N} v_{\mu }
\right) +N^2 E_C l_0 \ p_0 {1\over -\Delta} p_0 +
{p_i^2\over 2l_0E_J} \ .\
\label{pfjjc}
\end{eqnarray}

Following \cite{pol}\ we use the longitudinal part of the integer vector field
$v_{\mu }$ to shift the integration domain of $\Phi$. To this end we
decompose $v_{\mu }$ as follows:
\begin{equation}v_{\mu }= \Delta _{\mu } m + \Delta _{\mu }\alpha + K_{\mu \nu}
\psi _{\nu }\ ,
\label{dif}
\end{equation}
where $m\in Z$, $|\alpha |<1$ and $K_{\mu \nu }$ defined in (\ref{kbig}) . Here the
vectors $\psi _{\mu }$ are not integer, but they are nonetheless restricted
by the fact that the combinations $q_{\mu } \equiv \hat K_{\mu \nu }v_{\nu }=
\hat K_{\mu \alpha }K_{\alpha \nu } \psi_{\nu }$ must be integers. The
original sum over the three independent integers $\{ v_{\mu } \}$ can thus
be traded for a sum over the four integers $\{ m, q_{\mu } \}$ subject to
the constraint $\hat \Delta _{\mu }q_{\mu } =0$. The sum over the integers
$\{ m \}$ can then be used to shift the $\Phi$ integration domain from
$[-\pi /N, +\pi /N )$ to $(-\infty, +\infty )$. The integration over $\Phi$
is now trivial and enforces the constraint $\hat \Delta _{\mu }p_{\mu } =0$:
\begin{eqnarray}Z &&= \sum _{\{ q_{\mu } \}} \delta_{\hat \Delta_{\mu }
q_{\mu }, 0} \ \int {\cal D}p_{\mu } \ \delta \left(
\hat \Delta _{\mu }p_{\mu } \right) \ {\rm exp }(-S)\ ,\nonumber \\
S &&= \sum_x -i2\pi \ p_{\mu } K_{\mu \nu } \psi_{\nu } + N^2 l_0 E_C \ p_0
{1\over -\Delta } p_0 + {p_i^2 \over 2l_0E_J} \ .\
\label{pfjjd}
\end{eqnarray}

We now solve the two constraints by introducing a real gauge field $b_{\mu }$
and an integer gauge field $a_{\mu }$:
\begin{eqnarray} p_{\mu } &&\equiv K_{\mu \nu }b_{\nu } \ , \qquad
\qquad b_{\mu }\in R\ ,\nonumber \\
q_{\mu } &&\equiv \hat K_{\mu \nu } a_{\nu }\ ,\qquad
\qquad a_{\mu }\in Z\ .\
\label{sco}
\end{eqnarray}
By inserting the first of these two equations and by summing by parts, the
first term in the action (\ref{pfjjd}) \ reduces to $\sum_x -i2\pi b_{\mu }q_{\mu }$.
By inserting the second of the above equations and by summing by parts
again, this term of the action finally reduces to the mixed Chern-Simons
coupling $\sum_x -i2\pi \ a_{\mu }K_{\mu \nu }b_{\nu }$. Using the Poisson
formula (\ref{poi}) \ we can finally make $a_{\mu }$ also real at the expense of
introducing a set of integer link variables $\{ Q_{\mu }\}$ satisfying
the constraint $\hat \Delta_{\mu }Q_{\mu }$, which guarantees
gauge invariance:
\begin{eqnarray}Z &&= \sum _{\{ Q_{\mu } \}} \int {\cal D}a_{\mu }
\int {\cal D}b_{\mu } \ {\rm exp}(-S) \ ,\nonumber \\
S &&= \sum_x -i2\pi \ a_{\mu }K_{\mu \nu }b_{\nu } + N^2l_0E_C \ p_0 {1\over
-\Delta }p_0 + {p_i^2\over 2l_0E_J} + i2\pi a_{\mu }Q_{\mu } \ .\
\label{pfjje}
\end{eqnarray}
In this representation $K_{\mu \nu }b_{\nu }$ represents the conserved
three-current of charges, while $\hat K_{\mu \nu }a_{\nu }$ represents
the conserved three-current of vortices. Note that, actually, both these
conserved currents are integers (the factors of $N$ are explicit): indeed, the
summation over $\{ Q_{\mu } \}$ makes $a_{\mu }$ (and
therefore also $\hat K_{\mu \nu }a_{\nu }$) an integer, and then
the summation over $\{ a_{\mu }
\}$ makes $K_{\mu \nu}b_{\nu }$ an integer. The third term in the action
(\ref{pfjje}) \ contains two parts: the longitudinal
part $\left( p_i^L \right) ^2$ describes the
Josephson currents and represents a kinetic term for the charges;
the transverse part $\left( p_i^T \right) ^2$ can be rewritten as
a Coulomb interaction term for the vortex density $q_0$ by solving
the Gauss law enforced by the Lagrange multiplier $b_0$.

The partition function (\ref{pfjje}) \ displays
a high degree of symmetry between the charge and the vortex
degrees of freedom.
The only term which breaks this symmetry (apart from the integers $Q_{\mu }$)
is encoded in the kinetic term for the charges (Josephson currents).
This near-duality between charges and vortices has already been often invoked
in the literature \cite{jos} \ to explain the experimental quantum phase diagram
at very low temperatures. Here we introduce what we call the
{\it self-dual approximation} of Josephson junction arrays. This consists
in adding to the action in (\ref{pfjje}) \ a bare kinetic term for the vortices
(note that such a kinetic term is anyhow induced by
integrating out the charge degrees of freedom) and combining this with
the Coulomb term for the charges into $\sum_x {\pi ^2 \over N^2 l_0
E_C} q_i^2$. The coefficient is chosen so that the transverse
part of this term reproduces exactly the Coulomb term for the charges
upon solving the Gauss law enforced by the Lagrange multiplier $a_0$. The
longitudinal part, instead, represents the additional bare kinetic term
for the vortices. Given that now the gauge field $a_{\mu }$ has acquired
a kinetic term, we are also forced to introduce new integers $M_{\mu }$
via the Poisson formula to guarantee that the charge current $K_{\mu \nu }
b_{\nu }$ remains an integer:
\begin{eqnarray}Z_{SD} &&= \sum_{\{ Q_{\mu } \} \atop \{ M_{\mu } \} }
\int {\cal D}a_{\mu } \int {\cal D}b_{\mu } \ {\rm exp}(-S_{SD})\ ,\nonumber \\
S_{SD} &&= \sum_x -i2\pi \ a_{\mu }K_{\mu \nu }b_{\nu } + {p_i^2\over 2l_0E_J}
+{\pi ^2 q_i^2 \over N^2l_0E_C} + i2\pi a_{\mu }Q_{\mu } +i 2\pi b_{\mu }
M_{\mu } \ ,\
\label{sda}
\end{eqnarray}
where the new integers satisfy the constraint $\hat \Delta _{\mu }M_{\mu }=0$
to guarantee gauge invariance.
After a rescaling
\begin{eqnarray}A_0 &&\equiv {2\pi \over \sqrt{N}l_0}a_0\ ,\qquad \qquad
A_i \equiv {2\pi \over \sqrt{N}l} a_i \ ,\nonumber \\
B_0 &&\equiv {2\pi \over \sqrt{N}l_0} b_0 \ ,\qquad \qquad
B_i \equiv {2\pi \over \sqrt{N}l} b_i \ ,\
\label{res}
\end{eqnarray}
we obtain finally
\begin{eqnarray}Z_{SD} &&= \sum_{ \{ Q_{\mu } \} \atop \{ M_{\mu } \} }
\int {\cal D}A_{\mu } \int {\cal D}B_{\mu } \ {\rm exp} (-S_{SD}) \ ,\nonumber \\
S_{SD} &&= \sum_x {l_0 l^2\over 2e^2} \ F_iF_i -i {l_0l^2 \kappa \over 2\pi }
A_{\mu }k_{\mu \nu }B_{\nu } + {l_0l^2 \over 2g^2} \ f_if_i \nonumber \\
&&\ \ \ \ \ \ \ \ + i\sqrt{\kappa}
\left( l_0Q_0A_0 + lQ_iA_i \right) +i \sqrt{\kappa} \left( l_0M_0B_0 +
lM_iB_i \right) \ ,\
\label{sdb}
\end{eqnarray}
where $F_i$ and $f_i$ are defined in (\ref{ldfs}) \ and
\begin{equation}e^2 = 2N E_C\ ,\qquad \kappa = N \ ,
\qquad g^2 = {4\pi ^2 \over N }E_J \ .
\label{ide}
\end{equation}
This is exactly the partition function of our lattice Chern-Simons model
(\ref{lpf}) \ in the limit of infinite magnetic permeability $\eta = \infty $ and
with additional, integer-valued link variables $Q_{\mu }$ and $M_{\mu }$
coupled to the two gauge fields. Note that, with the above identifications,
the topological Chern-Simons mass (\ref{tma}) \ coincides with the Josephson plasma
frequency:
\begin{equation}m=\sqrt{2N^2 E_CE_J} \ .
\label{tmjpf}
\end{equation}
In the physical case $N=2$ this reduces to $m=\sqrt{8E_CE_J}$.
{}From the kinetic terms in (\ref{sdb}) \ we can also read off the charge
and vortex masses:
\begin{eqnarray}m_q &&= {1\over l^2 g^2} = {N\over 4\pi ^2 l^2 E_J} \ ,\nonumber \\
m_{\phi } &&= {1\over l^2e^2} = {1\over 2Nl^2E_C} \ .\
\label{cvma}
\end{eqnarray}
In the regime $ml\le O(1)$, which is typically experimentally relevant, we can
choose $l_0=l$: in this case the infinite magnetic permeability constitutes
the only non-relativistic effect in the physics of Josephson junction arrays
in the self-dual approximation. However, we expect this
non-relativistic effect
to be irrelevant as far as the phase structure and the charge-vorticity
assignements are concerned. Therefore, for simplicity, we shall henceforth
consider the relativistic model, by setting $l_0=l$ and $\eta =1$, although
it is not hard to incorporate a generic value of $\eta $ into our
subsequent formalism:
\begin{eqnarray}Z_{SD} &&= \sum _{ \{ Q_{\mu } \} \atop \{ M_{\mu } \} }
\int {\cal D}A_{\mu } \int {\cal D}B_{\mu } \ {\rm exp} (-S_{SD}) \ ,\nonumber \\
S_{SD} &&= \sum_x {l^3\over 2e^2} F_{\mu }F_{\mu } -i{l^3\kappa \over 2\pi }
A_{\mu }k_{\mu \nu }B_{\nu } + {l^3\over 2g^2} f_{\mu }f_{\mu }
+il\sqrt{\kappa } A_{\mu }Q_{\mu } +il\sqrt{\kappa }
B_{\mu }M_{\mu } \ .\
\label{frm}
\end{eqnarray}
Josephson junction arrays in the self-dual approximation constitute thus a
further, experimentally accessible example of the ideas presented in \cite{froa}
 and \cite{fwz}. 
The action in (\ref{frm}) \ provides in fact a pure gauge theory
representation of a model of interacting charges and vortices, represented
by the conserved currents
\begin{eqnarray}q_{\mu }^{\rm charge} &&\equiv {{\kappa }^{3\over 2}\over
2\pi } \ k_{\mu \nu }B_{\nu } \ ,\nonumber \\
\phi _{\mu }^{\rm vortex} &&\equiv {1\over 2\pi {\kappa}^{1\over 2}}
\ \hat k_{\mu \nu }A_{\nu } \ ,\
\label{cav}
\end{eqnarray}
where the prefactors are chosen so that the quantum of charge is given by
$\kappa $, while the quantum of vorticity is given by $1/\kappa $ (factors
of $q_e$ and $2\pi $ are absorbed in the definitions of the gauge fields and
the coupling constants).

In this framework, the mixed Chern-Simons term represents both the Lorentz
force caused by vortices on charges (coupling of $q_{\mu }^{\rm charge}$ to
the "electric" gauge field $A_{\mu }$) and, by a summation by parts,
the Magnus force 
caused by charges on vortices (coupling of $\phi _{\mu }
^{\rm vortex}$ to the "magnetic" gauge field $B_{\mu }$).
The integer-valued link variables $Q_{\mu }$ and $M_{\mu }$ represent
the (Euclidean) {\it topological excitations} \cite{pol} \ in the model.
They satisfy the constraints
\begin{eqnarray}\hat d_{\mu } Q_{\mu } &&=0\ ,\nonumber \\
\hat d_{\mu } M_{\mu } &&= 0\ .\
\label{const}
\end{eqnarray}
In a dilute phase they constitute
closed electric ($Q_{\mu }$) and magnetic ($M_{\mu }$) loops
on the lattice; in a dense phase there is the additional possibility of
infinitely long strings. Due to the constraints (\ref{const}) \ we can choose to
represent these topological excitations as
\begin{eqnarray}Q_{\mu } &&\equiv lk_{\mu \nu }Y_{\nu }\ ,
\qquad \qquad Y_{\nu }\in Z\ ,\nonumber \\
M_{\mu } &&\equiv l\hat k_{\mu \nu } X_{\nu } \ ,\qquad \qquad
X_{\mu } \in Z\ ,\
\label{frep}
\end{eqnarray}
and reabsorb them in the mixed Chern-Simons term as follows:
\begin{equation}S_{SD}=\sum _x {\dots } -i{l^3\kappa \over 2\pi} \left( A_{\mu }
-{2\pi \over l\sqrt{\kappa }} X_{\mu } \right) k_{\mu \nu } \left( B_{\nu }
- {2\pi \over l\sqrt{\kappa }} Y_{\mu } \right) + {\dots } \ .
\label{reabb}
\end{equation}
In this representation it is clear that the topological excitations render
the charge-vortex coupling {\it periodic} under the shifts
\begin{eqnarray}A_{\mu } &&\to A_{\mu } + {2\pi \over l\sqrt{\kappa }}
\ a_{\mu }\ ,\qquad \qquad a_{\mu } \in Z\ ,\nonumber \\
B_{\mu } &&\to B_{\mu } + {2\pi \over l\sqrt{\kappa }} \ b_{\mu }\ ,\qquad
\qquad b_{\mu } \in Z\ .\
\label{shif}
\end{eqnarray}
In physical terms, the topological excitations implement the well-known
\cite{fra}
\ periodicity of the charge dynamics under the addition of an integer
multiple of the flux quantum $1/\kappa $ per plaquette and
the (less-known) periodicity of the vortex dynamics under the addition of
an integer multiple of the charge quantum $\kappa $ per site.

If we would require that the full action (including charge-charge and
vortex-vortex interactions) (\ref{frm}) \ be periodic under the shifts (\ref{shif}) , then
we would obtain the compact Chern-Simons model studied in \cite{dst} . In this case
the relevant topological excitations would be essentially $iX_{\mu }$
and $iY_{\mu }$: since these can also describe finite open strings, there
is the additional possibility of electric and magnetic monopoles \cite{pol} .
As we showed in \cite{dst}, these monopoles play a crucial role in
the regime $ml\ll 1$.

\section{Phase structure analysis}
In this section we investigate symmetry aspects and non-perturbative features
of the model (\ref{frm})  due to the periodicity of the charge-vortex interactions
encoded in the mixed Chern-Simons term. As expected, these depend entirely on
the topological excitations which enforce the periodicity.

Upon a Gaussian integration the partition function (\ref{frm}) factorizes
readily as
\begin{equation} Z_{SD} = Z_{CS} \cdot Z_{\rm Top} \ ,
\label{fact}
\end{equation}
where $Z_{CS}$ is the pure gauge part defined in (\ref{lpf}) and
\begin{eqnarray}Z_{\rm Top} &&= \sum_{ \{ Q_{\mu } \} \atop \{ M_{\mu } \} }
\ {\rm exp}\left( -S_{\rm Top} \right) \ ,\nonumber \\
S_{\rm Top} &&= \sum_x {e^2\kappa \over 2l} \ Q_{\mu }{\delta _{\mu \nu }
\over {m^2- \nabla ^2 }} Q_{\nu } + {g^2 \kappa \over 2l} \ M_{\mu }
{\delta _{\mu \nu } \over {m^2-\nabla ^2}} M_{\nu } \nonumber \\
&&\ \ \ \ \ \ \ \ +i {2\pi m^2 \over l} \ Q_{\mu } {k_{\mu \nu }\over
{\nabla ^2
\left( m^2- \nabla ^2 \right) }} M_{\nu } \ ,\
\label{top}
\end{eqnarray}
with $m$ defined in (\ref{tma}), describes the contribution due to the topological
excitations. The phase structure of our model is thus determined by the
statistical mechanics of a coupled gas of closed or infinitely long electric
and magnetic strings with short-range Yukawa interactions. The scale $(1/m)$
represents the width of these strings. In our case it is of the order of the
lattice spacing $l$. The third term
in the action (\ref{top}), describing the topological Aharonov-Bohm interaction of
electric and magnetic strings, vanishes for strings separated by
distances much
bigger than $(1/m)$: in this case the denominator reduces to
$m^2\nabla ^2$ and, by using either one of the two equations in (\ref{frep}) and
the constraints
(\ref{const}) one recognizes immediately
that the whole term in the action  reduces to ($i2\pi\ {\rm integer}$),
which is
equivalent to $0$ (this reflects the fact that the original
charges and vortices satisfy the Dirac quantization condition).

\subsection{Free energy arguments}
In order to establish the phase diagram of our model we use the free energy
arguments for strings introduced in \cite{kog}.

The usual argument for strings with Coulomb interactions \cite{kog}
is that interactions between strings are unimportant
for the phase structure because small strings interact via short-range dipole
interactions, while large strings have most of their multipole
moments canceled
by fluctuations. This argument is even stronger in our case, where the
interaction is anyway short-range. Therefore one retains only the self-energy
of strings, which is proportional to their length, and phase transitions from
dilute to dense phases appear when the entropy of large strings, also
proportional to their length, overwhelms the self-energy.
We shall also neglect the interaction term between electric and magnetic
strings (imaginary term in the action (\ref{top}) ). This is clearly a good
approximation if both types of topological excitations are dilute.

Thus, one assigns a free energy
\begin{equation}F = \left( {le^2\kappa \over 2}G(ml) \ Q^2 + {lg^2\kappa \over 2}
G(ml) M^2 - \mu  \right) \ N
\label{free}
\end{equation}
to a string of length $L=lN$ carrying electric and magnetic quantum numbers
$Q$ and $M$, respectively. Here $G(ml)$ is the diagonal element of the lattice
kernel $G(x-y)$ representing the inverse of the operator $l^2 \left( m^2 -
\nabla ^2 \right) $. Clearly $G(ml)$ is a function of the dimensionless
parameter $ml$. The last term in (\ref{free}) represents the entropy of the string:
the parameter $\mu $ is given roughly by $\mu = {\rm ln} 5$, since at each
step the string can choose between 5 different directions. In (\ref{free}) we have
neglected all subdominant functions of $N$, like a ${\rm ln } N$
correction to the entropy.

The condition for condensation of topological excitations is obtained by
minimizing the free energy (\ref{free}) as a function of $N$. If the coefficient
of $N$ in (\ref{free}) is positive, the minimum is obtained for $N=0$ and
topological excitations are suppressed. If, instead, the same coefficient
is negative, the minimum is obtained for $N= \infty $ and the system will
favour the formation of large closed loops and infinitely long strings.
Topological excitations with quantum numbers $Q$ and $M$ condense therefore
if
\begin{equation}{le^2\kappa G(ml)\over 2\mu } \ Q^2 +
{lg^2 \kappa G(ml)\over 2\mu } \ M^2 < 1\ .
\label{cco}
\end{equation}
If two or more condensations are allowed by this condition one has to choose
the one with the lowest free energy.

The condition (\ref{cco}) describes the interior of an ellipse with semi-axes
$2\mu /(le^2\kappa G(ml))$ and $2\mu /(lg^2\kappa G(ml))$ on a square lattice
of integer electric and magnetic charges. The phase diagram is obtained by
investigating which points of the integer lattice lie inside the ellipse
as its semi-axes are varied. We find it convenient to present the results in
terms of the dimensionless parameters $lm$ and $e/g$:
\begin{eqnarray}{ml G(ml) \pi \over \mu } < 1 &&\to \cases{{e\over g}<1\ ,
& electric condensation\ ,\cr {e\over g} >1\ , & magnetic condensation \ ,\cr }
\nonumber \\
{ml G(ml) \pi \over \mu } > 1 &&\to  \cases{{e\over g} <
{\mu \over ml G(ml) \pi} \ , & electric condensation\ ,\cr
{\mu \over ml G(ml) \pi }<{e\over g}<{ml G(ml) \pi \over \mu } \ ,
& no condensation\ ,\cr
{e\over g} > {ml G(ml) \pi \over \mu }\ , & magnetic condensation \ .\cr 
 }\
\label{phc}
\end{eqnarray}
As expected, these condensation patterns are symmetric around the
the point $e/g =1$, reflecting the self-duality of the model.
In first approximation the electric (magnetic) condensation phase is
characterized by the fact that $\{ Q_{\mu } \}$ ($\{ M_{\mu } \}$) fluctuate
freely, while all $M_{\mu }=0$ ($Q_{\mu }=0$). Within this approximation it is
clearly consistent to neglect altogether the interaction term between
electric and magnetic strings in (\ref{free}). Taking into account small loop
corrections in the various phases can lead to a renormalization of
coupling constants and masses and, correspondingly, to a shift of the
critical couplings $(ml)_{\rm crit}$ and $(e/g)_{\rm crit}$
for the phase transitions. A notable exception is the case
in which there is only one phase transition: in this case the
critical coupling
is $(e/g)_{\rm crit}=1$ due to self-duality.

\subsection{Wilson and 't Hooft loops}
In order to distinguish the various phases we introduce the typical
order parameters of lattice gauge theories \cite{pol}, namely the {\it Wilson loop} for
an electric charge $q$ and the {\it 't Hooft loop} for a vortex $\phi $:
\begin{eqnarray}L_W &&\equiv {\rm exp} \left( i {q\over {\kappa }^{1\over 2}}
\sum_x lq_{\mu }A_{\mu } \right) \ ,\nonumber \\
L_H &&\equiv {\rm exp} \left( i \phi {\kappa }^{3\over 2}
\sum_x l\phi _{\mu }B_{\mu }\right) \ ,\
\label{opa}
\end{eqnarray}
where $q_{\mu }$ and $\phi _{\mu }$ vanish everywhere but on the links of the
closed loops, where they take the value 1. Since the loops are closed
they satisfy
\begin{equation}\hat d_{\mu }q_{\mu }=\hat d_{\mu }\phi _{\mu } =0 \ .
\label{cloo}
\end{equation}

The expectation values $\langle L_W \rangle $ and $\langle L_H \rangle $ can
be used to characterize the various phases. First of all they
measure the interaction potential between static, external
test charges $q$ and $-q$ and vortices $\phi $ and $-\phi $, respectively
\cite{pol} . Secondly, by representing the closed loops $q_{\mu }$ and $\phi _{\mu }$
as
\begin{eqnarray}q_{\mu } &&\equiv lk_{\mu \nu }A_{\nu }^q\ ,\nonumber \\
\phi _{\mu } &&\equiv l \hat k_{\mu \nu }A_{\nu }^{\phi } \ ,\
\label{lst}
\end{eqnarray}
we can rewrite the Wilson and 't Hooft loops as
\begin{eqnarray}L_W &&= {\rm exp} \left( i{q\over {\kappa }^{1\over 2}} \sum_x
l^2 A_{\mu }^q F_{\mu } \right) \ ,\nonumber \\
L_H &&= {\rm exp} \left( i {\kappa }^{3\over 2} \phi  \sum_x
l^2 A_{\mu }^{\phi }f_{\mu } \right) \ ,\
\label{whl}
\end{eqnarray}
which is a lattice version of Stoke's theorem, the integers $A_{\mu }^q$ and
$A_{\mu }^{\phi }$ ($=0,\pm 1$) representing the area elements of the surfaces
spanned by the closed loops.
The second terms of the expansions of $\langle L_W \rangle $ and $\langle
L_H \rangle $ in powers of $q$ and $\phi $ measure therefore the gauge
invariant correlation functions $\langle F_{\mu }(x) F_{\nu }(y) \rangle $ and
$\langle f_{\mu }(x) f_{\nu }(y) \rangle $.
Third, if we represent $\phi _{\mu }$ as
\begin{equation}\phi \phi _{\mu } \equiv {l^2 \over 2\pi } \ \hat k_{\mu \nu }
A_{\nu }^{\rm e.m.}\ ,
\label{nint}
\end{equation}
we can also rewrite the 't Hooft loop as
\begin{equation}L_H = {\rm exp} \left( i\sum_x l^3
A_{\mu }^{\rm e.m.}q_{\mu }^{\rm charge} \right) \ .
\label{elc}
\end{equation}
With the interpretation of $A_{\mu }^{\rm e.m.}$ as an
external electromagnetic
gauge potential the expectation value of the 't Hooft loop measures the
{\it electromagnetic response} of the system in the various phases.
An analogous relation clearly holds for the Wilson loop.

The expectation values of the Wilson and 't Hooft loops are easily
obtained by combining the definitions \ref{opa} \ with \ref{frm}:
\begin{eqnarray}\langle L_W \rangle &&= {{Z_{\rm Top} \left( Q_{\mu } +
{q\over \kappa } q_{\mu }, M_{\mu } \right) } \over {Z_{\rm Top} \left(
Q_{\mu }, M_{\mu } \right) }} \ ,\nonumber \\
\langle L_H \rangle &&= {{Z_{\rm Top } \left( Q_{\mu }, M_{\mu }+ \phi \kappa
\phi _{\mu } \right) } \over {Z_{\rm Top} \left( Q_{\mu }, M_{\mu } \right)
}} \ ,\
\label{evwh}
\end{eqnarray}
where the notation is self-explanatory. In the following we shall analyze
these expressions in the various phases obtained in (\ref{phc}). We shall mostly
only indicate the form of small loop corrections: a full renormalization
group analysis is beyond the scope of the present paper and we won't be
able to predict the orders of the phase transitions.

Let us begin with the {\it electric condensation phase}. In this phase the
ground state contains many infinitely long electric strings $Q_{\mu }$.
These have a crucial effect on the gauge symmetry associated with the gauge
field $A_{\mu }$. To see this let us consider a gauge transformation
$A_{\mu } \to A_{\mu }+ d_{\mu } \Lambda $, where, for simplicity, we take
$\Lambda $ as a function of the component $x^1$ only. If we choose the usual
boundary conditions $F_{\mu }=f_{\mu }=0$ at infinity, the change of
the action
(\ref{frm}) under the above gauge transformation is given by
\begin{equation}\Delta S_{SD}= \sum_{x^0, x^2} i\sqrt{\kappa } \left(
\Lambda (x^1=+\infty ) Q_1(x^1=+\infty )- \Lambda (x^1=-\infty ) Q_1(x^1=
-\infty ) \right) \ .
\label{cha}
\end{equation}
In a dilute phase, with only small closed loops, $Q_1(x^1=+\infty )=Q_1(x^1=
-\infty ) =0$ and the action is automatically gauge invariant. In a dense
phase, with many infinitely long strings, $Q_1(x^1=+\infty )$ and $Q_1(x^1=-
\infty )$ are generically different from zero. Gauge invariance requires that
$\Delta S_{SD}$ vanishes modulo $i2\pi $. In the dense phase this is realized
only if $\Lambda $ takes the values
\begin{equation}\Lambda = {2\pi \over \sqrt{\kappa }} n\ , \qquad \qquad n\in Z\  ,
\label{asb}
\end{equation}
at infinity. This means that, in the electric condensation phase, the
{\it global} gauge symmetry is spontaneously broken down to the discrete
gauge group
$Z$, so that the total (global) symmetry of this phase is $Z_A \times R_B$.

The Wilson loop expectation value takes a particularly simple form if the
external test charges are multiples of the charge quantum: $q=n\kappa $, $n
\in Z$. In fact, since we sum over $\{ Q_{\mu } \} $, the integer loop
variables $nq_{\mu }$ can be absorbed by a redefinition of the appropriate
$Q_{\mu }$'s, with the result
\begin{equation}\langle L_W(q=n\kappa ) \rangle = 1\ .
\label{pscr}
\end{equation}
This indicates that, in this phase, external test charges $q=n\kappa $ are
perfectly {\it screened} by the topological excitations and behave
thus freely.
In order to compute the Wilson loop expectation value for generic $q$
we have to
perform explicitly the sum over $\{ Q_{\mu } \}$. To this end we have to
remember the constraint $\hat d_{\mu }Q_{\mu }=0$. We solve this constraint
by representing $Q_{\mu }=lk_{\mu \nu }n_{\nu }$ and summing over $\{ n_{\mu }
\} $, with the appropriate gauge fixing. We then use Poisson's formula \ref{poi}
\ to turn this sum into an integral, by introducing a new set of integer
link variables $\{ k_{\mu } \}$ satisfying $\hat d_{\mu }k_{\mu }=0$ in order
to guarantee the gauge invariance under $n_{\mu }\to n_{\mu }+ld_{\mu } i$.
At this point we can perform explicitly the Gaussian integration over
$\{ n_{\mu } \}$. In the approximation of neglecting terms proportional to
$\nabla ^2/m^2$ (keeping such terms would not alter substantially the result)
the new integers $\{ k_{\mu } \}$ can be absorbed by a redefinition of the
magnetic topological excitations $\{ M_{\mu } \}$, giving the result:
\begin{eqnarray}\langle L_W \rangle &&= {{Z_{\rm corr.} \left( q_{\mu }
\right) }\over {Z_{\rm corr.}\left( q_{\mu }=0 \right)}} \ ,\nonumber \\
Z_{\rm corr.}\left( q_{\mu } \right) &&= \sum _{\{ M_{\mu } \} {\rm loops}}
{\rm exp} \sum_x \left( -{g^2\kappa \over 2l} \ M_{\mu } {{\delta _{\mu \nu }}
\over -\nabla ^2} M_{\nu } +i 2\pi {q\over \kappa} A^q_{\mu }M_{\mu } \right)
\ .\
\label{corrwil}
\end{eqnarray}
Since the magnetic topological excitations are in a dilute phase we have
to sum only over small closed loops: in this phase the dominant part
of ${\rm ln} \langle L_W \rangle$ vanishes for generic $q$ and the whole
result is given by small loop corrections. These are identical in form
to the small loop corrections for the correlation functions in the
low-temperature phase of the three-dimensional XY model; correspondingly
the Wilson loop expectation value can be computed by exactly the same
low-temperature expansion used for the XY model. The first-order term
in this expansion is obtained by considering only the smallest possible
lattice loops and gives the result
\begin{equation}\langle L_W \rangle = {\rm exp} \left( 2 {\rm e}^{-{g^2 \kappa l
\over 6}} \sum_{x, \mu } \left[ {\rm cos}
\left( 2\pi {q\over \kappa } q_{\mu }
\right) -1\right] \right) \ .
\label{perwillo}
\end{equation}
The periodicity of this result is a direct consequence of the spontaneous
symmetry breaking $R_A \to Z_A$. This implies also that the gauge invariant
correlation function reduces to
\begin{equation}\langle F_{\mu }(x) F_{\nu }(y) \rangle \propto
\left( \delta _{\mu \nu } \nabla ^2 - d_{\mu }\hat d_{\nu } \right)
\ {\delta _{x,y}\over l^3} \ ,
\label{ecpapcf}
\end{equation}
which is essentially a contact term on the scale of the lattice spacing.

The computation of the 't Hooft loop expectation value follows exactly the
same lines as the above computation of the Wilson loop. The results is
\begin{eqnarray}\langle L_H \rangle &&= {\rm exp}
\left( -{g^2\kappa ^3 \phi ^2\over 2l}
\sum_x \phi_{\mu } {\delta _{\mu \nu }\over -\nabla ^2}\phi_{\nu } \right)
\ {{Z_{\rm corr}\left( \phi_{\mu }\right) }\over Z_{\rm corr} \left(
\phi_{\mu }=0\right) }\ ,\nonumber \\
Z_{\rm corr}\left( \phi_{\mu } \right) &&= \sum_{\{ M_{\mu } \} {\rm loops}}
{\rm exp} \left( -{g^2\kappa \over 2l} \sum_x  M_{\mu }
{\delta _{\mu \nu } \over -\nabla ^2} M_{\nu }
+2\kappa \phi \ M_{\mu } {\delta _{\mu \nu } \over -\nabla ^2}
\phi_{\nu } \right) \ .\
\label{hla}
\end{eqnarray}
The first few terms in the expansion of the small loop corrections can again
be computed with the same techniques as in the low-temperature phase of the
XY model.
One finds that their contribution amounts to perturbative corrections of
the Coulomb coupling constant $g^2\kappa ^3\phi ^2/2l$ of the
dominant term in (\ref{hla}).

>From (\ref{hla}) we can extract the nature of the electric condensation phase.
First of all, by considering, as usual, a rectangular loop of length $T$ in
the imaginary time direction and of length $R$ in one of the spatial
directions and computing the dominant large-$T$ behaviour of ${\rm ln}
\langle L_H\rangle $ we find that the interaction potential between external
test vortices of strength $\phi $ and $-\phi $ is proportional to ${\rm ln}R$.
Vortices are thus logarithmically {\it confined}, which amounts to the
{\it Meissner effect}. Secondly, by using the representations (\ref{lst}) and
\ref{whl} , we find the correlation function
\begin{equation}\langle f_{\mu }(x) f_{\nu }(y) \rangle \propto
{{\delta _{\mu \nu } \nabla ^2 -d_{\mu }\hat d_{\nu }} \over \nabla ^2}
\ {{\delta_{x,y}}\over l^3}\ ,
\label{lrcf}
\end{equation}
which is long-range, indicating that the "$B_{\mu }$-photon"
is {\it massless}.
This is the massless excitation associated with the spontaneous symmetry
breaking of the global gauge symmetry $R_A \to Z_A$.
Third, by using the representations (\ref{nint}) and (\ref{elc}), we find that the
induced electromagnetic current is given by
\begin{equation}J_{\mu }^{\rm e.m.} \propto \left( \delta_{\mu \nu }-
{{d_{\mu }\hat d_{\nu }}\over  \nabla ^2} \right) \ A_{\nu }^{\rm e.m.}\ ,
\label{iemc}
\end{equation}
which is the standard {\it London form}. We thus conclude that the electric
condensation phase is actually a {\it superconducting phase}.

No further computation is needed to extract the nature of the {\it magnetic
condensation phase}: this is the exact dual of the electric condensation
phase just described. Specifically, the global gauge symmetry associated
with $B_{\mu }$ is spontaneoulsy broken down to $Z_B$, so that the total
symmetry of this phase is $R_A \times Z_B$. Correspondingly, the
"$A_{\mu }$-photon" is {\it massless} and the $\langle F_{\mu }(x)F_{\nu }
(y) \rangle $ correlation function is long-range. Electric charges are
logarithmically
{\it confined}, which means that an infinite energy (voltage) is required
to separate a neutral pair of charges. We call this phase with infinite
resistance a {\it superinsulator}. In real Josephson junction arrays we
expect however the conduction gap to be large but finite due to the small
ground capacity $C_0$, resulting in a normal insulator.

If $mlG(ml)\pi /\mu >1$ a third phase can open up between the superconducting
and superinsulating phases. In this third phase both the electric and the
magnetic topological excitations are dilute. Far away from the
phase transitions and to first approximation we can neglect them altogether.
This gives the result
\begin{eqnarray}\langle L_W \rangle &&= {\rm exp}
\left( -{e^2q^2\over 2l\kappa } \ q_{\mu }
{{\delta _{\mu \nu }}\over {m^2-\nabla ^2}}q_{\nu } \right) \ ,\nonumber \\
\langle L_H \rangle &&= {\rm exp} \left( -{g^2 \phi ^2 \kappa ^3\over 2l}
\ \phi_{\mu } {{\delta _{\mu \nu }}\over {m^2-\nabla ^2}}
\phi_{\nu } \right) \ .\
\label{whmp}
\end{eqnarray}
Small loop corrections to these results can be obtained by restricting the
$\{ Q_{\mu } \}$ and $\{ M_{\mu } \}$ sums in (\ref{evwh}) to small closed loops
and using again the same techniques as in the low-temperature expansion
of the XY model. These will lead to perturbative corrections of the
coupling constants and masses in (\ref{whmp}); however the first-order result
(\ref{whmp}) is enough to establish the nature of this phase. The global
symmetry characterizing this phase is $R_A\times R_B$ and, corrrespondingly,
both "photons" are massive, resulting in short-range correlation
functions $\langle F_{\mu }(x)F_{\nu }(y)\rangle$ and $\langle f_{\mu }(x)
f_{\nu }(y)\rangle $. Both charges and vortices interact via short-range
Yukawa potentials and behave thus freely when separated by distances larger
then the scale $(1/m)$. In presence of any dissipation mechanism (which would
not alter the other two phases) this third phase corresponds thus to
a {\it metallic} phase of the Josephson junction array.

In conclusion we can represent the phase diagram of our model as follows:
\begin{eqnarray}{ml G(ml) \pi \over \mu } < 1 &&\to \cases{{e\over g}<1\ ,
& superconductor ($Z_\times R_B$)\ ,\cr {e\over g} >1\ ,
& superinsulator ($R_A\times Z_B$)\ ,\cr } \nonumber \\
{ml G(ml) \pi \over \mu } > 1 &&\to
\cases{{e\over g} < {\mu \over ml G(ml) \pi} \ ,
& superconductor ($Z_A\times R_B$)\ ,\cr
{\mu \over ml G(ml) \pi }<{e\over g}<{ml G(ml) \pi \over \mu } \ ,
& metal ($R_A\times R_B$)\ ,\cr
{e\over g} > {ml G(ml) \pi \over \mu }\ , & superinsulator ($R_A\times Z_B$)
\ ,\cr } \
\label{fphc}
\end{eqnarray}
where we have indicated in parenthesis the global symmetries of the various
phases. 

A numerical computation of the function $mlG(ml)\pi /\mu $ for the
value $\mu ={\rm ln}5$, gives an indication that a window for
the metallic
phase is open for $ml$ just larger than 1, while in the regime $ml \le O(1)$,
relevant for Josephson junction arrays, a single phase transition from a
superconductor to a superinsulator at $(e/g) =1$ is favoured.

The experimental results for Josephson junction arrays, which are essentially
resistance measurements as a function of temperature in arrays with
O($10^4$) cells, shows, etrapolating at  zero-temperature,
a quantum phase transition between an insulator and a superconductor
in the vicinity of the self-dual point $E_J/E_C= 2/\pi ^2 \simeq 0.2$. This is in perfect agreement
with $(e/g) =1$. In fact, we have: $E_J/E_C= (2 g^2)/(e^2 \pi^2)$ and, for $(e/g) =1$, we get
$E_J/E_C= 2/\pi ^2 \simeq 0.2$.

\section{Superconductor with topological order}

Topologically ordered superconductors have a long-distance hydrodynamic action which 
can be entirely formulated in terms of generalized compact gauge fields, the dominant 
term being the topological BF action.

BF theories are  topological theories that can be defined on manifolds $M_{d+1}$ of any 
dimension (here d is the number of spatial dimensions) and play a crucial role in models of two-dimensional gravity
\cite{marito}.
In \cite{dst} we have shown that the BF term also plays a crucial role in the physics of Josephson junction arrays.
 
The BF term \cite{birmi} is the wedge product of a p-form B  and the curvature $d A$ of a (d-p) form A:

\begin{equation}
S_{BF} = {k \over 2 \pi} \int_{M_{d+1}}  B_p \wedge d A_{d-p}\ ,
\nonumber
\end{equation}
where $k$ is a dimensionless coupling constant.
This can also be written as
\begin{equation}
S_{BF} = {k \over 2 \pi} \int_{M_{d+1}}  A_{d-p} \wedge d B_p\ .
\label{sbf1}
\end{equation}
The integration by parts does not imply any surface term since we will concentrate on compact spatial manifolds without
boundaries and we require that the fields go to pure gauge configuration at infinity in the time direction.
Indeed this action has a generalized Abelian gauge symmetry under the transformation 
\begin{equation}
B \rightarrow B + \eta \ ,
\nonumber
\end{equation}
where $\eta$ is a closed p form: $d \eta = 0$.
Gauge transformations:

\begin{equation}
A \rightarrow A + \xi \ ,
\nonumber
\end{equation}
with $\xi$ a closed (d-p) form instead, change the action by a surface term. This, however vanishes with the boundary conditions we have
chosen.

Here we will be interested in the special case where $A_1$ is a 1-form and, correspondingly, $B_{d-1}$ is a (d-1)-form: 
\begin{equation}
S_{BF} = {k \over 2 \pi} \int_{M_{d+1}} A_1 \wedge d B_{d-1} \ .
\label{sbf}
\end{equation}
In the special case of (3+1) dimensions,  B is the well-known Kalb-Ramond tensor field $B_{\mu\nu}$ \cite{kalb}. 

In the application to superconductivity, the conserved current $j_1 = * dB_{d-1}$ 
represents the charge fluctuations, while the generalized current 
$j_{d-1} = * dA_1$ describes the conserved fluctuations of (d-2)-dimensional 
vortex lines. As a consequence,
the form $B_{d-1}$ must be considered as a pseudo-tensor, while $A_1$ is a vector, as
usual. The BF coupling is thus P- and T-invariant. 

The low-energy effective theory of the superconductor can be entirely expressed in terms
of the generalized gauge fields $A_1$ and $B_{d-1}$. The dominant term at long distances is
the BF term; the next terms in the derivative expansion of the effective theory are the
kinetic terms for the two gauge fields (for simplicity of 
presentation we shall assume relativistic invariance), giving:

\begin{eqnarray}
S_{TM} &=  \int_{M_{d+1}} { - 1 \over 2 e^2} d A_1 \wedge * d A_1 +  {k \over 2 \pi} A_1 \wedge d B_{d-1} \nonumber \\
&+ {(- 1)^{d-1} \over 2 g^2} d B_{d-1}\wedge * d B_{d-1}\ ,
\label{topmas}
\end{eqnarray}  
where $e^2$ and $g^2$ are coupling constants of dimension $m^{-d+3}$ and $m^{d-1}$ respectively.

The BF-term is the generalization to any number of dimensions of the Chern-Simons mechanism for the topological mass
\cite{jac}. To see this let us now compute the equation of motion for the two forms $A$ and $B$:

\begin{equation}
{1 \over g^2} d*d B_{d-1} = {k \over 2 \pi} d A_1 \ ,
\label{eqmo1}
\end{equation} 
and
\begin{equation}
{1 \over  e^2} d*d A_1 = {k \over 2 \pi} d B_{d-1} \ .
\label{eqmo2}
\end{equation}  
Applying $d*$ on both sides of (\ref{eqmo1}) and (\ref{eqmo2}) we obtain 

\begin{eqnarray}
 &d*d* d A_1 - {k e^2\over 2 \pi} d*d B_{d-1} = 0 \ , \nonumber \\
 &d*d* d B_{d-1} - {k g^2\over 2 \pi} d*d A_1 = 0 \ .
\label{is}
\end{eqnarray} 
The expression $*d*$ is proportional  to $\delta$, the adjoint of the exterior derivative \cite{eh}.
Substituing  $d*d B_{d-1}$  and 
$d*d A_1$ in (\ref{is}) with the expression coming from (\ref{eqmo1}) and (\ref{eqmo2}) we obtain

\begin{eqnarray}
&\left( \Delta + m^2 \right) d A_1 = 0 \ , \nonumber \\
&\left( \Delta + m^2 \right) d B_{d-1} = 0 \ ,
\label{topmas1}
\end{eqnarray}                                                                                                                       
where $\Delta = d \delta$ (when acting on an exact form) and  $m = {k e g \over 2 \pi}$ is the topological mass.
This topological mass plays the role of the gap characterizing  the superconducting ground state.
Note that the gap arises here from a topological mechanism and not from a local order parameter acquiring a vacuum expectation value.
Equations (\ref{eqmo1}) and (\ref{eqmo2}) tell us that charges are sources for vortex 
line currents encircling them and viceversa. This is the coupling between charges and
vortices at the origin of the gap. A related mechanism for topological mass generation in (3+1)-dimensional gauge theories is the
generalization of the Schwinger mechanism proposed in \cite{dvali}.

Let us now consider the special case of (2+1) dimensions (d=2). In this case also
$B$ becomes a (pseudo-vector) 1-form and, correspondingly the BF term reduces to a
mixed Chern-Simons term. This can be diagonalized by a transformation $A={a+b\over 2}$, 
$B=a-b$, giving
\begin{equation}
S_{BF}(d=2) = {k\over 4\pi} \ \int a \wedge da - {k\over 4\pi} \ \int b \wedge db \ .
\label{newa}
\end{equation}
The result is a doubled Chern-Simons model for gauge fields of opposite chirality. This action, including its non-Abelian generalization with
kinetic terms was first considered in \cite{roman}. It is
the simplest example of the class of P- and T-invariant topological phases of strongly
correlated (2+1)-dimensional electron systems considered in \cite{freedman}. Indeed, the BF
term is the natural generalization of such doubled Chern-Simons models to any dimension.
Doubled (or mixed) Chern-Simons models are thus particular examples in two spatial
dimensions of a wider class of P- and T-invariant topological fluids that have a 
superconducting phase. These fluids are described by the topological BF theory
with compact support for both gauge fields.

Topological BF models provide also a generalization of anyons to arbitrary dimensions. While in (2+1) dimensions fractional statistics arises
from the representations of the braid group, encoding the exchange of particles, in (3+1) dimensions it arises from the adiabatic transport of
particles around vortex strings and, in  (d+1) dimensions, from the motion of an hypersurface $\Sigma_h$ around  another hypersurface $\Sigma_{d-h}$.
The relevant group in this case is the motion group and the statistical parameter is given by ${2 \pi \over k} h (d-h)$, where
$k$ is the BF coupling constant \cite{szab}.

Let us now illustrate the mechanism of superconductivity. To this end we shall from now on  consider only rational $k = {k_1 \over k_2}$ with
$k_i$ integers, and specialize to manifolds $M_{d+1} = M_d \times R_1$, whith $R_1$ representing the time direction. 

The compactness of the gauge fields allows for the presence of topological defects,
both electric and magnetic. The electric topological defects  couple to the form
$A_1$ and are string-like objects  described by a singular closed 1-form $Q_1$. Magnetic topological 
defects couple to the form
$B_{d-1}$ and are closed (d-1)-branes described  by a singular (d-1)-dimensional form $\Omega_{d-1}$. These forms
represent the singular parts of the field strenghts $dA_1$ and $dB_{d-1}$, allowed by the compactness of the gauge
symmetries \cite{pol}, and are such that the  integral of their Hodge dual  over any  hypersurface of dimensions d and 2, respectively, is $2 \pi$ times
an integer as can be easily derived using a lattice regularization. Contrary to the currents $j_1$ and $j_{d-1}$, which represent charge- and
vortex-density waves, the topological defects describe localized charges and vortices. In the effective theory these have structure on the
scale of the ultraviolet cutoff.

We will not discuss here the conditions for the condensation of topological defects, but we will show, instead that the phase 
of electric condensation describes  a superconducting phase in any dimension.
A detailed analysis would require the use of an ultraviolet regularization.
Here we will present a formal derivation implying the 
ultraviolet regularization  (e.g. a lattice regularization).

In the phase in which electric topological defects condense (while magnetic ones are dilute) the partition function requires 
a formal sum also over the form $Q_1$

\begin{eqnarray}
&Z = \int {\cal D}A {\cal D}B {\cal D}Q \nonumber \\ 
&\exp \left[ i {k \over 2 \pi}\int_{M_{d+1}} \left( A_1 \wedge d B_{d-1} + A_1 \wedge *Q_1 \right) \right] \ .
\label{pfbf}
\end{eqnarray}
Let us now compute the expectation value of the 't Hooft  operator, 
$\langle L_H \rangle$, which represents
the amplitude for creating and separating a pair of vortices with fluxes $\pm \phi$:

\begin{eqnarray}
&\langle L_H \rangle = {1 \over Z} \int {\cal D}A {\cal D}B {\cal D}Q \nonumber \\ 
&\exp \left[ i {k \over 2 \pi}\int_{M_{d+1}} \left( A_1 \wedge d B_{d-1} + A_1 \wedge *Q_1 \right) \right. \nonumber \\
&+ \left. i {k \over 2 \pi} \phi \int_{S_{d-1}} B_{d-1} \right]\ .
\label{pfbh}
\end{eqnarray}
Using Stokes' theorem we can rewrite

\begin{equation}
\int_{S_{d-1}} B_{d-1}  = \int_{S_d} d B_{d-1} \ ,
\label{sto}
\end{equation}
where the surface $S_d$ is such that $\partial S_d \equiv  S_{d-1}$ and represents a compact orientable surface on $M_d$.
Inserting (\ref{sto}) in (\ref{pfbh}) and integrating over the field $A$ we obtain:

\begin{eqnarray}
\langle L_H \rangle  &  \propto \int  {\cal D}B {\cal D}Q \ \delta \left( d B_{d-1} + *Q_1 \right) \nonumber \\ 
&\exp \left[  i{k \over 2 \pi}  \phi \int_{S_d} d B_{d-1} \right]\ .
\end{eqnarray}
Integrating over B gives then:

\begin{equation}
\langle L_H \rangle = \propto \int {\cal D}Q\ \exp \left[ - i{k \over 2 \pi} \phi \int_{S_d} *Q_1 \right] \ .
\label{vev}
\end{equation}
The Poisson summation formula implies finally that the 't Hooft loop expectation value
vanishes for all flux strengths $\phi $ different from 
\begin{equation}
{\phi \over k_2} = {2\pi \over k_1} \ n \qquad \qquad n \in N \ .
\label{newb}
\end{equation}
This is nothing else than the Meissner effect, illustrating that the electric condensation
phase is superconducting. Indeed, the electric condensate carries $k_1$ fundamental
charges of unit $1/k_2$ as is evident from (\ref{pfbf}), and
correspondingly vortices must carry an integer multiple of the fundamental fluxon
$2\pi /({k_1/k_2})$. All other vorticities are confined: in this purely 
topological long-distance theory the confining force is infinite; including 
the higher order kinetic terms (\ref{topmas}) and the UV cutoff one would recover a generalized area law. 

Another way to see this is to compute the current induced by an external electromagnetic
field $A_{\rm ext}$. The corresponding coupling is $\int_{M_{d+1}} A_{\rm ext}
\wedge \left( * j_1 + *Q_1 \right)$ $\propto $ $\int_{M_{d+1}} A_{\rm ext}
\wedge \left( dB_{d-1} + *Q_1 \right)$. Since $A_{\rm ext}$ can be entirely reabsorbed in a redefinition
of the gauge field $A_1$, the induced current vanishes identically, $j_{\rm ind} =0$.
This is just the London equation in the limit of zero penetration depth. Including the
higher-order kinetic terms for the gauge fields and the UV cutoff one would again recover the standard form
of the London equation. 

Associated with the confinement of vortices there is a breakdown of the original U(1)
matter symmetry under transformations $A_1 \to A_1 + d\lambda$. To see this let us consider 
the effect of such a transformation on the partition function (\ref{pfbf}) with an
electric condensate. Upon integration by parts, the exponential of the action acquires 
a multiplicative factor
\begin{equation}
{\rm exp} \ i {k_1\over 2\pi k_2} \left( \int_{M_d , t=+\infty} \lambda \wedge * Q_1
-\int_{M_d , t=-\infty} \lambda \wedge * Q_1 \right) \ .
\label{newc}
\end{equation}
Assuming a constant $\lambda $, we see that the only values for which the partition
function remains invariant are 
\begin{equation}
\lambda = 2\pi \ n \ {k_2 \over k_1}\ , \qquad \qquad n=1 \dots k_1 \ ,
\label{newd}
\end{equation}
which shows that the global symmetry is broken from U(1) to $Z_{k_1}$. Note that this
is not the usual Landau mechanism ofspontaneous symmetry breaking. Indeed, there is
no local order parameter and the order is characterized rather by 
the expectation value of non-local, topological operators.

The hallmark of topological order is the degeneracy of the ground state on manifolds with non-trivial topology as shown by Wen \cite{wen1}. 
In (2+1) dimensions the degeneracy for the mixed Chern-Simons term was proven in \cite{hoso} for the case of integer coefficient $k$ 
of the Chern-Simons term.

The degeneracy of the ground state of the BF theory on a manifold with non-trivial topology  was proven in \cite{gord} in (3+1) dimensions.
This result can be generalized to compact topological BF models in any number of dimensions \cite{szab}.
Consider the model (\ref{sbf1}) with $k = {k_1 \over k_2}$ on a manifold $M_d \times R_1$, 
with $M_d$ a compact, path-connected , orientable d-dimensional manifold without boundaries. The degeneracy of the ground state is expressed
in terms of the intersection matrix $M_{mn}$ \cite{bott} with $m,n = 1....N_p$ and $N_p$  the rank of the matrix, 
between  p-cycles and  (d-p)-cycles. $N_p$ corresponds to the number of generators of the two homology groups
$H_p(M_d)$ and $H_{d-p}(M_d)$ and is essentially the number of non-trivial cycles on 
the manifold $M_d$. The degeneracy of the ground state is given by $|k_1 k_2 M|^{N_p}$,
where $M$ is the integer-valued determinant of the linking matrix. In our case p = (d-1) and the degeneracy reduces to

\begin{equation}
|k_1 k_2 M|^{N_{d-1}} \ .
\label{deged}
\end{equation}

\section{Frustration}

The gauge theory formulation of JJA \cite{dst} clearly shows
that the superconducting ground state is a P- and T- invariant
generalization of Laughlin's incompressible quantum fluid. The
simplest example of a topological fluid \cite{fwz} is a ground state
described by a low energy effective action given solely by the
topological Chern-Simons term $$S = k/4\pi\ \int
d^3x \ A_{\mu} \epsilon^{\mu \nu \alpha}
\partial_{\nu} A_{\alpha}$$ for a compact $U(1)$ gauge field
$A_{\mu}$, whose dual field strength $F^{\mu} = \epsilon^{\mu \nu
\alpha} \partial_{\nu} A_{\alpha}$ yields the conserved matter
current. In this case the degeneracy of the ground state on a
manifold of genus $g$ will be $k^g$ (or $(k_1 k_2)^g$ if $k =
k_1/k_2$ is a  rational number). For planar unfrustrated JJA one
finds that the topological fluid is described by two $k=1$
Chern-Simons gauge fields of opposite chirality and, thus, there is
no degeneracy of the ground state \cite{dst}, \cite{bff}.

In \cite{bff} we argued that frustrated JJA may support a
topologically ordered ground state with non-trivial degeneracy on higher genus
surfaces, described by a pertinent
superconducting quantum fluid, thus providing a more interesting 
example of a system in which superconductivity arises from
the topological mechanism proposed in \cite{nsm} rather than from
the usual Landau-Ginzburg mechanism. The role of frustration in the
determination of the ground state degeneracy of JJA (in the limit $E_J \gg E_C$)
was also analyzed in \cite{ds}.

The fact that frustration can change the coefficient of the Chern-Simons term
can be easily understood by the following example. Let us briefly review the mean
field approach to spin liquid states \cite{aff}. This description is based on the
Hubbard model that, following Anderson, is thought to be relevant  for high $T_c$
superconductivity. At half filling this model reduces to the Heisenberg model.
Although very little is known rigourosly about the Heisenberg model on a square
lattice, it is believed that the Heisenberg antiferromagnet is N\'eel ordered at
T = 0.  Is it possible to drive the model toward a disordered ground state to 
give rise to a
spin liquid? The answer to this question is yes, and the key element to reach
disorder is frustration. An example is to consider next to nearest neighbor
interactions in the Heisenberg model: the result is a chiral spin liquid 
that breaks P- and
T-invariance  \cite{wwz}. We will thus concentrate on the Heisenberg model, taking the point
of view that the various spin liquid phases can be realized with small modification of its
Hamiltonian \cite{fra}. Moreover our analysis can be exactly repeted for the
Heisenberg model with next to nearest neighbour interactions presented in
\cite{wwz}.

We will study a mean field theory for the Heisenberg antiferromagnet
originally proposed in\cite{aff}.
To this end we consider a
system of spin $S = 1/2$ on a square lattice with nearest neighbor interactions
and periodic boundary conditions. The
Hamiltonian that describes this model is: 
\begin{equation}H = \sum_{\langle ij \rangle} J_{ij} {\bf S_i}{\bf S_j} \ .
\label{hei}
\end{equation} 
To obtain the mean field ground state for the spin liquid we use a fermion
representation of the spin operator:
\begin{equation}
{\bf S_i} = {1 \over 2}c{^+}_{i\alpha} {\bf \sigma_{\alpha \beta}} c_{i\beta} \ ,
\label{sf}
\end{equation}
where ${\bf \sigma}$ are the Pauli matrices. The operators (\ref{sf}) reproduce
the spin 1/2 algebra of  only if $c{^+}_{i\alpha}c_{i\alpha} =  1$, namely if
there is only one fermion per site (half filling).
Substituting (\ref{sf}) in (\ref{hei}) we obtain (up to an additive constant):
\begin{equation}H = {1 \over 4} \sum_{\langle ij \rangle} J_{ij}c{^+}_{i\alpha}
c_{j\alpha}c{^+}_{j\beta}c_{i\beta} \ .
\label{ff}
\end{equation}
Introducing the Hubbard-Stratonovich field $\chi_{ij}$ we can rewrite the
four-fermion interaction as (we will concentrate here only on this term):
\begin{equation}H = ...- \sum_{\langle ij \rangle}c{^+}_{i\alpha}\chi_{ij}
c_{j\alpha} \ .
\label{tf}
\end{equation}
We will now treat the field $\chi_{ij}$ within the mean field
 approximation $\chi_{ij}^{\rm MF}$
and, since it is complex, we will parametrize it with an amplitude $\rho_{ij}$
and a phase $A_{ij}$. Self consistency implies $\chi_{ij}^{\rm MF} = \langle c{^+}_{i\alpha}
c_{j\alpha} \rangle$.

Various possible mean-field solutions have been studied. We will be interested
in the case in which $\chi_{ij}^{\rm MF}$ generates a flux: the fermions behave
as though they would be moving in a magnetic field and the P- and T-symmetries are broken;
moreover, when an integer number of Landau levels is filled, the  electron gas
is incompressible.
For  the case in which the flux per plaquette is $\pi$,
(half of quantum of flux), it has been shown that, after integrating out the fermion fields, the effective
theory contains a Chern-Simons term with a coefficient $k =2$ \cite{wwz}(note that in the construction 
presented in \cite{wwz}
in a certain range of parameter P- and T-symmetry are broken even with a flux
$\pi$ per plaquette). This corresponds
to the case in which the first Landau level is completely filled. It also been
shown that for generalized flux phases, where the flux per plaquette is $2\pi
p/q$ with $q$ an even integer, the effective
theory contains a Chern-Simons term with a coefficient $k = q$ (for $q=2$ we
obtain the previous case, namely 1/2 quantum flux per plaquette). In this case we have
$q/2$ Landau levels that are filled.

Let us now consider the hamiltonian:
\begin{equation}H = \sum_{\langle ij \rangle} J_{ij} S_i U_{ij} S_j \ ,
\label{hef}
\end{equation} 
where $U_{ij}$ is a frustration field that satisfies : $\prod_p U_{ij} = \exp
2i\pi f$. Here $f$ is the frustration parameter. It acts as a fictitious
magnetic field, and, due to the invariance of the theory under $f \rightarrow f
+n, n \in Z$ and $f \rightarrow -f$ we have $0 \le f \le {1\over 2}$. Moreover
since we have imposed periodic boundary conditions the flux must be a rational
multiple of the quantum flux: $f = 2 \pi p/q$.
Let us follow the same line of reasoning as before. To this end with start with a flux $\pi$ per plaquette 
(first Landau level filled),
coming from the mean field approximation of (\ref{tf}). In this case we end up with a flux: $2 \pi (1/2 +
p/q) = 2 \pi (q +2p)/2q$ per plaquette. We will end up with the problem of the
hopping of fermions in a fictitius magnetic field that is changed from $\pi
\rightarrow 2 \pi (q +2p)/2q$, with $q$ Landau levels filled. This frustrated model will be described by a
Chern-Simons theory with coefficient $k = 2q$ while the unfrustrated
model corresponds to a Chern-Simons theory with coefficient $k = 2$.

Let us now go back to JJA. 
It has been suggested that, in presence of
$n_q$ offset charge quanta per site and $n_{\phi }$ external magnetic
flux quanta per plaquette in specific ratios,
Josephson junction arrays might have incompressible quantum fluid
\cite{Lau} \ phases corresponding to purely two-dimensional
{\it quantum Hall phases}
for either charges \cite{odi}  or vortices \cite{choi,as}.

In \cite{dst} we have shown that, 
if quantum Hall phases for charges or vortices
are realized, then JJA naturally support a topologically ordered
ground state and a phase in which they behave as a topological
superconductor \cite{nsm}; there is, in fact, a renormalization of
the Chern-Simons coefficient yielding a non-trivial ground state
degeneracy on the torus (and in general on manifolds with
non-trivial topology). To implement a torus topology we impose
doubly periodic conditions at the boundary of the square lattice.

To study this degeneracy let us consider the low energy limit
of the partition function (\ref{sda}).
For the following analysis we will take the time as a continuos parameter (we
remeber that the discretization of time is allowed by the fact that charges are
quantized, but is not a necessary condition). In
this limit we have:
\begin{eqnarray}
Z &&= \sum_{\{ Q_0 \} \atop \{ M_0 \} } \
\int {\cal D}A_{\mu } \int {\cal D}B_{\mu } \ {\rm exp}(-S)\ ,
\nonumber \\
S &&= \int dt \sum_{\bf x} -i{ 1\over 2\pi }\ A_{\mu }K_{\mu \nu }B_{\nu } + i
 A_0 Q_0 + i  B_0 M_0\ .
\label{ac}
\end{eqnarray}
This form of the partition function holds true also with toroidal
boundary conditions. With continuos time the operator $K_{\mu \nu}$ is
defined by $K_{00} = 0$, $K_{0i} = -\epsilon_{ij}
d_j$, $K_{i0} = S_i \epsilon_{ij} d_j$ and $K_{ij} = -S_i
\epsilon_{ij} \partial_0$, in terms of forward (backward) shift and
difference operators $S_i$ ($\hat S_i$) and $d_i$ ($\hat d_i$). Its
conjugate $\hat K_{\mu \nu}$ is defined by $\hat K_{00} = 0$, $\hat
K_{0i} = - \hat S_i \epsilon_{ij} \hat d_j$, $\hat K_{i0} =
\epsilon_{ij} \hat d_j$ and $\hat K_{ij} = - \hat S_j \epsilon_{ij}
\partial_0$.

The topological excitations are described by the integer-valued
fields $Q_0$ and $M_0$ and represent unit charges and vortices
rendering the gauge field components $A_0$ and $B_0$ integers via
the Poisson summation formula; their fluctuations determine the
phase diagram \cite{dst}. In the classical limit
the magnetic excitations are dilute and the charge excitations condense
rendering the system a superconductor: vortex confinement amounts
here to the Meissner effect. In the quantum limit, the magnetic
excitations condense while the charged ones become dilute: the
system exhibits insulating behavior due to vortex superconductivity
accompanied by a charge Meissner effect.

 By rewriting the topological excitations as the
curl of an integer-valued  field
\begin{eqnarray}Q_{0 } &\equiv K_{0 i }Y_{i }\ ,
\qquad \qquad Y_{i }\in Z\ ,\nonumber \\
M_{0 } &\equiv \hat K_{0 i } X_{i } \ ,\qquad \qquad
X_{i } \in Z\ ,
\label{fre}
\end{eqnarray}
we get the mixed Chern-Simons term as follows:

\begin{eqnarray}
Z &= \sum_{\{ X_i, Y_i \}} \
\int {\cal D}A_{\mu } \int {\cal D}B_{\mu } \ {\rm exp}\ (-S)\ ,
\nonumber \\
S &= -{1\over 2\pi }i \int dt \sum_{\bf x} A_0 K_{0i}
\left( B_i - 2 \pi Y_i  \right) \nonumber \\ &+ B_0 \hat K_{0i}
\left( A_i - 2 \pi X_i  \right)
+ A_i K_{ij} B_j \ . \label{ai}
\end{eqnarray}
>From (\ref{ai}) one sees that the gauge field components $A_i$ and
$B_i$ are angular variables due to their invariance under
time-independent integer shifts. Such shifts do not affect the last
term in the action, which contains a time derivative, and may be
reabsorbed in the topological excitations $X_i$ and $Y_i$, leaving
also the first term of the action invariant. The low energy theory
is thus compact.

In analogy with the conventional quantum Hall setting one should
expect the charge and vortex transport properties to depend on the
ratios of the offset charges (i.e. the filling fractions)
$(n_q/n_{\phi})$ and $(n_{\phi}/n_q)$, respectively. Due to the
 periodicity of the charge-vortex coupling, however, $n_{\phi }$ ($n_q$)
is defined only modulo an integer as far as charge (vortex)
transport properties are concerned. Using this freedom one may
define effective filling fractions (we shall assume $n_q \ge 0$,
$n_{\phi}\ge 0$ for simplicity) as
\begin{eqnarray}\nu _q &\equiv
{n_q \over {n_{\phi }-\left[ n_{\phi } \right] ^- +
\left[ n_q \right] ^+ }} \ , \qquad \qquad 0\le \nu_q\le 1\ ,\nonumber \\
\nu_{\phi } &\equiv {n_{\phi } \over {n_q-\left[ n_q \right] ^- +\left[
n_{\phi } \right] ^+}} \ ,\qquad \qquad 0\le \nu_{\phi }\le 1 \ ,
\label{eff}
\end{eqnarray}
where $\left[ n_q \right] ^{\pm }$ indicate the smallest (greatest)
integer greater (smaller) than $n_q$. Of course, these effective
filling fractions are always smaller than 1.

In \cite{dst} we assumed the existence of these quantum Hall phases
and discussed them in the framework of the gauge theory
representation of Josephson junction arrays, showing that, depending
on certain parameters of the array there are both a charge quantum
Hall phase and a vortex quantum Hall phase. Here we will concentrate
on the low energy limit of the charge quantum Hall phase and we will
show  that the system has topological order and behaves as a
superconductor when charge condenses.

The pertinent low energy theory is now given by:
\begin{equation} S = \int dt\sum_{\bf x} - {i \over \pi}
A_{\mu }K_{\mu  \nu }B_{\nu } - {i \nu _q
\over \pi }A_{\mu }K_{\mu  \nu }A_{\nu } \ ,
\label{csag}
\end{equation}
with $\nu_q =p/n$. The main difference with (\ref{ac}) \ is the
addition of a  pure Chern-Simons term for the $A_{\mu }$ gauge field
. We have also rescaled the coefficient of the mixed Chern-Simons
coupling by a factor of 2 (compare with (\ref{ac}) ). This factor of
2 is a well-known aspect of Chern-Simons gauge theories \cite{wil}.
Moreover, since in JJA the charge degrees of freedom are bosons,
the allowed \cite{dst} filling fractions are given by $
\nu_q={p\over n}$ , with $ pn = {\rm even \ integer}$ in accordance
with \cite{read}. As a result, the action (\ref{csag}) may now be
written in terms of two independent gauge fields $A_\mu$ and
$B_\mu^q = B_\mu + \nu_q A_\mu$ yielding:
\begin{equation} S = \int dt\sum_{\bf x} - {i \over \pi}
A_{\mu }K_{\mu  \nu }B_{\nu }^q \ .
\label{csnf}
\end{equation}

In describing JJA one has to require the periodicity of
charge-vortex couplings; the coupling of the topological excitations
enforcing the periodicity of the mixed Chern-Simons term $A_{\mu
}K_{\mu \nu }B_{\nu }^q$ is then:
\begin{equation}
S = \int dt \sum _x \dots + ip A_{0 }
Q_{0 } + in B_{0 }M_{0 } \ ,
\label{topex}
\end{equation}
that can be rewritten as:
\begin{equation}
S = \sum _x \dots + ilp A_{0 } \left( Q_{0 }+M_{0 } \right) + iln
B^q_{0 }M_{0} \ .\label{ntope}
\end{equation}
Due to the replacement $B_{\mu }\to B_{\mu }^q$, the periodicities
of the two original gauge fields are
\begin{eqnarray}A_{i } &\to A_{i } + \pi n
\ a_{i } \ , \qquad \qquad a_{i } \in Z\ ,\nonumber \\
B_{i } &\to B_{i } + \pi p \ b_{i } \ ,\qquad
\qquad b_{i }\in Z \ ,
\label{newshiq}
\end{eqnarray}
and
\begin{equation}
B^q_{i } \to B^q_{i } + \pi p \ b_{i } \ ,\qquad
\qquad b_{i }\in Z \ .
\label{bper}
\end{equation}
The resulting low energy theory is thus, again, compact.

Using the representation (\ref{fre}), one may rewrite the mixed
Chern-Simons term as
\begin{eqnarray}S = \int dt  \sum_x {\dots } &-i{(pq /2)\over 2\pi}
 \left({2 A_{0 }\over n}
\right) K_{0 i} \left({2 B^q_{i }\over p}
- 2\pi Y_{i } \right) \nonumber \\
 &-i{(pq /2)\over 2\pi}{2 B^q_{0 }\over n}
 K_{0 i } \left({2 A_{i}\over n}
- 2\pi X_{i } \right) \ .
\label{reab}
\end{eqnarray}
In this representation it is clear that the topological excitations render
the charge-vortex coupling  periodic under the shifts
\begin{eqnarray}A^{'}_{i } = {2 A_{i }\over n} &\to A^{'}_{i } + 2\pi
\ a_{i }\ ,\qquad \qquad a_{i } \in Z\ , \nonumber \\
B^{'}_{i } = {2 B^q_{i}\over p} &\to B^{'}_{i } + 2\pi  \ b_{i }\ ,\qquad
\qquad b_{i } \in Z\ .
\label{shi}
\end{eqnarray}
This model corresponds to two Chern-Simons terms with coefficients $
\pm k/4 \pi$ with $k = n p/2$ an integer. It is worth to point out
that, since $B_\nu^q$ does not have a definite parity (is a liner
combination of a vector and a pseudovector) the model is not P- and
T-invariant, as it must be due to the presence  of the Chern-Simons
term for the field $A_\mu$.

The hallmark of topological order is the degeneracy of the ground
state on manifolds with non-trivial topology 
\cite{wen}. The torus degeneracy on the lattice of the Chern-Simons
model was computed in \cite{elsem}. For a single Chern-Simons term
this degeneracy is $(k)^g$ where $k$ is the integer coefficient of
the Chern-Simons term, and $g$ the genus of the surface. In our case
this degeneracy is $2 \times (k)^g = 2 \times {n p \over 2}$, since
we have two Chern-Simons terms. This degeneracy is exactly what is
expected for a doubled Chern-Simons model \cite{freedman}, for which
the physical Hilbert space is the direct product of the two Hilbert
spaces of the component models.

We will now demonstrate that the phase where topological excitations $Q_0$
condense while $M_0$ are dilute describes an
effective gauge theory of a superconducting state. The partition function is:
\begin{eqnarray}
Z_{LE} &&= \sum_{\{ Q_0 \} } \
\int {\cal D}A_{\mu } \int {\cal D}B_{\mu }^q \ {\rm exp}\ (-S)\ ,
\nonumber \\
S &&=  \int dt \sum_{\bf x} -{i k \over 2 \pi }\ A_{\mu }K_{\mu \nu }B_{\nu }^q
+ {i k \over 2 \pi } A_0 (2\pi  Q_0)\ .
\label{af}
\end{eqnarray}

 To this end note
first that a unit external charge, represented by an additional term
$ i2\pi a_0 (t, {\bf x}) \delta_{{\bf x}{\bf x_0}}$  is completely
screened by the charge condensate, since it can be absorbed into a
redefinition of the topological excitations $Q_0$. In order to
characterize the superconducting phase we introduce the typical
order parameter namely the 't Hooft loop of length $T$ in the time
direction:
\begin{equation}
L_H \equiv {\rm exp} \left( i \phi {\kappa \over 2 \pi}
\int dt \sum_x \phi _{\mu }B_{\mu }\right) \ ,
\label{thlo}
\end{equation}
 where $\phi_0 (t, {\bf x})$ = $\left( \theta
\left( t+T/2 \right) - \theta \left( t-T/2 \right) \right)
\delta_{{\bf x} {\bf x_1}}$ - $\left( \theta \left( t+T/2 \right) -
\theta \left( t-T/2 \right) \right) \delta_{{\bf x} {\bf x_2}}$ and
$\phi_i (-T/2, {\bf x})$, $\phi_i (T/2, {\bf x})$ are unit links joining
${\bf x_1}$ to ${\bf x_2}$ and ${\bf x_2}$ to ${\bf x_1}$ at fixed
time and vanishing everywhere else. Its vacuum expectation value
$\langle L_H \rangle $ yields the amplitude for creating a separated
vortex-antivortex pair of flux $\phi $, which propagates for a time
$T$ and is then annihilated in the vacuum.

Since we replaced $B_{\mu } \to   B_{\mu }^q$, we may rewrite the 't
Hooft loop as:
\begin{equation}
L_H \equiv {\rm exp} \left( i \phi {\kappa \over 2 \pi}
\int dt \sum_x \left( \phi _{\mu }B_{\mu }^q - {p\over n}  A_\mu \phi_\mu \right)
\right) \ .
\label{nthlo}
\end{equation}

To compute $\langle L_H
\rangle $ one should integrate first over the gauge field $B_{\mu }^q$
to get
\begin{eqnarray}
\langle L_H \rangle &&\propto \sum_{\{ Q_0 \} } \ \int {\cal
D}A_{\mu } \  \delta \left( \hat K_{\mu \nu} A_{\nu} - \phi \phi_{\mu}
\right) \times
\nonumber \\
&& \ {\rm exp} i \phi {\kappa \over 2 \pi}\left( \int dt \sum_{\bf x}
 A_0 (2\pi  Q_0) - {p\over n}  A_\mu \phi_\mu
\right) \ .
\label{ah}
\end{eqnarray}
The sum over $Q_0$ enforces the condition that ${k\over 2 \pi}A_0$ be an integer.
As a consequence, $\hat K_{i 0} A_0 = {2 \pi \over k} n_i = \phi
\phi_i$ with $n_i$ an integer.
We thus have:
\begin{equation}\phi = {2 \pi \over k} q, q  \in Z \ ;
\label{fluq}
\end{equation}
thus, $\langle L_H \rangle $ vanishes for
all fluxes different from an integer multiple of the fundamental
fluxon, which is just the Meissner effect. In the low-energy
effective gauge theory vortex-antivortex pairs are confined by an
infinite force which becomes logarithmic upon including also
higher-order Maxwell terms.

>From (\ref{reab}) one has (depending if $p$ is an even integer or
$n$ is an even integer) either that $n$ is the charge unit with
$p/2$ units of charge, or, viceversa, that $p$ is the charge unit
with $n/2$ units of charge. By rewriting (\ref{fluq}) as $\phi = {2
\over pn} 2 \pi n$ one finds the standard flux quantization $\phi =
{2 \pi \over N e} n$ where $e$ is the charge unit and $N$ is the
number of units of charge.

In this paper we have derived a superconductivity mechanism which is not based on the 
usual Landau theory of spontaneous symmetry breaking. Our considerations here focused on 
the low-energy effective theory in order to expose
the physical basis of the topological superconducivity mechanism. It is however crucial
to stress that the simplest example (k=1) of 
this type of topological superconductivity is concretely realized as the global 
superconductivity mechanism in 
planar Josephson junction arrays, as we have shown in \cite{dst}. Naturally, it
would be most interesting to find examples of microscopic models realizing this
superconductivity mechanism with more complex degeneracy patterns.
A possibilty is represented by models with frustration.
The frustrated planar JJA we have analized provide, in the quantum Hall phases,
an explicit example of both topological order with non-trivial ground state degeneracy 
on manifolds with non-trivial topology, and of a new
superconducting behavior \cite{bff} analogous to Laughlin's quantum
Hall fluids.

\end{document}